\UseRawInputEncoding
\documentclass[conference]{IEEEtran}
\IEEEoverridecommandlockouts
\usepackage{cite}
\usepackage{amsmath,amssymb,amsfonts}
\usepackage{algorithmic}
\usepackage{graphicx}
\usepackage{textcomp}
\usepackage{xcolor}
\usepackage{siunitx}
\def\BibTeX{{\rm B\kern-.05em{\sc i\kern-.025em b}\kern-.08em
    T\kern-.1667em\lower.7ex\hbox{E}\kern-.125emX}}
\begin{document}

\title{An application of reinforcement learning to residential energy storage under real-time pricing\\}

\author{\IEEEauthorblockN{Eli Brock}
\IEEEauthorblockA{\textit{Electrical and Computer Engineering} \\
\textit{University of Pittsburgh}\\
Pittsburgh, USA \\
etb28@pitt.edu}
\and
\IEEEauthorblockN{Lauren Bruckstein}
\IEEEauthorblockA{\textit{Computer Science} \\
\textit{University of Pittsburgh}\\
Pittsburgh, USA \\
leb127@pitt.edu}
\and
\IEEEauthorblockN{Patrick Connor}
\IEEEauthorblockA{\textit{Electrical and Computer Engineering} \\
\textit{University of Pittsburgh}\\
Pittsburgh, USA \\
pmc40@pitt.edu}
\and
\IEEEauthorblockN{Sabrina Nguyen}
\IEEEauthorblockA{\textit{Electrical and Computer Engineering} \\
\textit{University of Pittsburgh}\\
Pittsburgh, USA \\
san86@pitt.edu}
\and
\IEEEauthorblockN{Robert Kerestes}
\IEEEauthorblockA{\textit{Electrical and Computer Engineering} \\
\textit{University of Pittsburgh}\\
Pittsburgh, USA \\
rjk39@pitt.edu}
\and
\IEEEauthorblockN{Mai Abdelhakim}
\IEEEauthorblockA{\textit{Electrical and Computer Engineering} \\
\textit{University of Pittsburgh}\\
Pittsburgh, USA \\
maia@pitt.edu}
}

\maketitle

\begin{abstract}
With the proliferation of advanced metering infrastructure (AMI), more real-time data is available to electric utilities and consumers. Such high volumes of data facilitate innovative electricity rate structures beyond flat-rate and time-of-use (TOU) tariffs. One such innovation is real-time pricing (RTP), in which the wholesale market-clearing price is passed directly to the consumer on an hour-by-hour basis. While rare, RTP exists in parts of the United States and has been observed to reduce electric bills. Although these reductions are largely incidental, RTP may represent an opportunity for large-scale peak shaving, demand response, and economic efficiency when paired with intelligent control systems. Algorithms controlling flexible loads and energy storage have been deployed for demand response elsewhere in the literature, but few studies have investigated these algorithms in an RTP environment. If properly optimized, the dynamic between RTP and intelligent control has the potential to counteract the unwelcome spikes and dips of demand driven by growing penetration of distributed renewable generation and electric vehicles (EV). This paper presents a simple reinforcement learning (RL) application for optimal battery control subject to an RTP signal.
\end{abstract}


\section{Introduction}

\subsection{Real-Time Pricing}
Conventionally,  consumers pay a flat rate for electricity, set by the local supplier months in advance. Despite the flat-rate market structure's popularity, its lack of sensitivity to the wholesale market creates inefficiencies.

Electricity consumption varies dramatically over the course of a day, with spikes in the morning and early evening and lulls in the nighttime. As consumption varies, so does wholesale price, driven by the law of demand as well as supply-side factors. The first marginal levels of electricity are met by the suppliers with the lowest marginal costs, typically renewable and non-dispatchable sources such as wind and solar. As more energy is demanded but is not available from these cheaper sources, the market shifts to more expensive suppliers, like nuclear, and eventually on to fossil fuels \cite{Woo2016Merit-orderMarkets, He2013ModelingGermany}.

Given the increasing marginal costs of supplying electricity, the economically optimal shape of the aggregate load curve would be a flat line. Consistent consumption levels reduce the average cost of electricity and drive down load peaks \cite{Deng2015AApproaches}. Lower peaks mitigate the need for highly expensive \textit{peaker plants}, or generation plants which only supply power during peak demand. In addition, generation equipment ratings can be lower, reducing capital costs.

Flat electricity rates hide these market dynamics from consumers who would otherwise be incentivized to shift their consumption habits, thereby counterbalancing the inefficient stratification of consumption. With the adoption of metering infrastructure (AMI) in highly industrialized nations, utilities can gather accurate information on individual customers' consumption levels at short intervals \cite{RashedMohassel2014AInfrastructure}. Naturally, new rate structures have emerged to take advantage of the demand response opportunities presented by AMI.

Among the most common of these progressive rate structures is time-of-use (TOU) pricing, which is an option for many customers in the US \cite{Wang2015Time-of-useUtilities,Celebi2012Time-of-useStructures}. TOU assigns different prices to different hours of the day, with high prices during typical peak hours and vice versa. Under such a structure, there could be two, three, or more pricing levels.

TOU, while an effective demand response tool, simply presents a less pervasive version of the flat-rate problem \cite{Celebi2012Time-of-useStructures}. Prices within on-peak and off-peak periods still vary, and the demand curve's shape varies from day to day and season to season. Real-time pricing (RTP), also known as \textit{hourly} or \textit{dynamic} pricing, goes one step further, with prices changing hour-by-hour subject to variable market conditions.

RTP is popular among economists who see it as a natural way to facilitate demand response and market efficiency \cite{Zethmayr2018ThePrices, Borenstein2005ThePricing, Allcott2011RethinkingPricing}. RTP has been sporadically deployed for large industrial customers in utilities across the US \cite{Nezamoddini2017Real-timeStates}. For residential customers, the primary examples are in Illinois, where the Public Utilities Act, passed in the early 2000's, mandated utilities provide RTP as an option.

There are different types of RTP programs. Some programs, such the Power Smart Pricing program offered by Illinois' Ameren utility, set their prices 24 hours before they take effect. These prices are based on the Transmission System Operator's (TSO) day-ahead prices. For such programs, RTP is perhaps a misnomer. True RTP programs, such as Illinois' Commonwealth Edison (ComEd) utility's Hourly Pricing\footnote{https://hourlypricing.comed.com/live-prices/}, charge consumers according to actual clearing prices. Each hour's price is the average of that hour's 12 five-minute wholesale clearing prices as listed by the Pennsylvania-Jersey-Maryland (PJM) interconnection, the regional TSO. Note that this price is not finalized until \textit{after} the hour has passed and the electricity has been consumed. In this sense, the Hourly Pricing program distinguishes itself from so-called hour-ahead RTP programs, where the price is forecasted one hour ahead. As such, it exhibits absolute fidelity to the wholesale market and presents perhaps the most challenging rate structure for smart demand response algorithms.

The literature shows promise for RTP \cite{Deng2015AApproaches}. ComEd's optional Hourly Pricing program had saved residential customers an average of 12\% on their electric bills from the program's inception in 2007 through 2011 \cite{Wang2011LessonsPrograms}. In \cite{Zethmayr2018ThePrices}, authors demonstrated that 97\% of ComEd customers \textit{would} save more money (an average of 13\% annual savings) under RTP relative to flat-rate without behavioral changes. 

Similar benefits exist for industrial customers. A 2017 investigation found that opting in to an available RTP program was financially advantageous for a majority of industrial case studies and was profitable for a larger range of industrial customers than TOU \cite{Nezamoddini2017Real-timeStates}. 

Most of the studies mentioned above analyze RTP by projecting an RTP signal onto consumers' existing load curves. Such studies are limited in that they do not account for potential changes in an RTP customer's behavior. The resultant savings, then, are incidental: they arise naturally from the flatter shape of certain individual consumer load curves relative to the grid at large.

Some papers have investigated consumer response to real-time prices \cite{Allcott2009RealMarkets, Allcott2011RethinkingPricing}. There is evidence consumers do indeed respond to price signals. However, since the studies are necessarily limited to customers who choose to opt-in to such programs, they are hardly representative. It is unlikely that the typical residential electricity consumer would consistently check real-time prices and adjust their usage habits or control EV or home storage systems. This is the motivation for automated, intelligent demand response for agents in RTP environments.

\subsection{Demand Response and Reinforcement Learning}
Typical price-driven demand response methods rely on mathematical optimization strategies, such as convex optimization and dynamic programming, to schedule electricity consumption and storage according to a price signal known well in advance (such as TOU) \cite{Deng2015AApproaches}. RTP introduces a new level of complexity, since the price of electricity is unknown in the immediate future and sometimes - as is the case with ComEd's Hourly Pricing - in the present.

Some researchers have employed price-prediction algorithms to handle RTP demand response \cite{Ren2011HomePrediction, Leon-garcia2010PriceElectricity, Erdinc2015SmartUtilization}. They use a mathematical algorithm to forecast the price signal then proceed with the mathematical optimization methods as they do with pre-scheduled prices. These are \textit{model-based} control strategies. By contrast, \textit{model-free} approaches arrive at control decisions directly from observations of the environment, without attempting to understand the underlying nature of the system (in this case, the future prices). Such problems can be expressed as Markov Decision Processes and solved using model-free reinforcement learning (RL) algorithms  \cite{Deng2015AApproaches}.

RL has been deployed for demand response applications. Reference \cite{Lu2018AApproach} uses RL to set optimal prices by playing out the economics between service providers and customers. The work in \cite{Wan2018Model-FreeLearning} subjects an EV to real-time prices and uses RL to find an optimal discharging strategy. Several other papers use similar model-free approaches to demand response, many involving EV charging schedules \cite{Wen2015OptimalLearning, Vandael2015ReinforcementMarket, Lu2018AGrid, Chis2017ReinforcementPrice, Ruelens2017ResidentialLearning}. However, to the best of the authors' knowledge, RL has not been applied to price-driven residential demand response in an actualized commercial RTP environment, such as ComEd's Hourly Pricing.

This paper uses the Deep Q Network (DQN) algorithm, introduced in \cite{MnihPlayingLearning}, as it is ideal for problems with discrete action spaces. DQN is based on Q-learning, an RL method in which an agent assigns a ``Q-value'' to each of its available actions during each time step. The Q-value of an action is the expected reward the agent will achieve given that it takes the action and behaves optimally afterwards. Higher Q-values represent higher-quality actions. In contrast to the original Q-learning algorithm, the DQN variant uses a deep neural network as a function approximator to learn the Q-function mapping state-action pairs to Q-values. The inputs to the neural network are the agent's observations of the environment. The output layer has one node per available action, representing the Q-value for each. A well-known DQN learning process has the agent take a random action with probability $\epsilon$, otherwise choosing the action with the highest predicted Q-value. This is called an \textit{epsilon-greedy} policy. It is designed to strike a balance between exploration and exploitation. After each step, the neural network is updated by gradient descent \cite{MnihPlayingLearning}.

\subsection{Demonstration}
Price-driven demand response takes many forms, including storage control and smart scheduling of dispatchable loads. In this paper, we consider a battery which is free to charge and discharge without holding energy in reserve for any particular purpose, such as powering an EV. An example of such a battery would be the Tesla Powerwall. The Powerwall already features some profit-maximizing functionality: consumers under TOU can provide the Powerwall with on- and off-peak hours and it will charge and discharge accordingly.

This paper presents a simple simulation of how such a battery, controlled by a DQN agent, may respond to an RTP signal. The battery operates as if it were a participant in the ComEd Hourly Pricing program. The objective is to charge and discharge such as to maximizes profit; the optimal policy will be a buy-low sell-high strategy.

The purpose of the demonstration is to investigate the potential of RL for optimal demand response in an RTP environment. The paper aims to evaluate the extent to which an RL algorithm learns to make profitable decisions and how well its strategy generalizes to unseen price signals. The following section will detail the proposed approach and outline the RL environment. Next, the method of implementation will be explained. The paper will then evaluate the performance of the agent in training and testing. The conclusion highlights the primary takeaways, followed by an exploration of future work based on the preliminary results.

\section{Proposed Approach: RL-Controlled Energy Storage}

This demonstration uses a simplified battery model. The battery can store a certain amount of energy, and can charge or discharge a fixed amount every hour. It is operating under ComEd's Hourly Pricing program, meaning it sees the price of electricity for an hour only after the hour has passed. This is the price it pays to charge as well the price it earns for discharging. The latter dynamic assumes net metering; that is, even if the discharged energy cannot be used to offset other parts of the load, it can simply be sold back to the grid at market value. Net metering is typical for customers with generation resources such as rooftop solar.

The battery makes its decision - charge, discharge, or idle - at the beginning of each hour and maintains that behavior during the entire hour. The battery can attempt to discharge when empty or charge when full, but the charge level will not change. Decisions are governed by a DQN agent.

The RL environment consists of four parameters, allowing the researcher to customize the price/battery environment:
\begin{enumerate}
    \item \textit{Price signal} ($\{p_0,p_1,\ldots,p_N\}$): $p_n$ represents the most recent electricity price observable during hour $n$ of the simulation. Note that this price is only operative during hour $n-1$. $N$ determines the length of an episode.
    \item \textit{Maximum charge} ($W$): Represents the maximum amount of energy the battery can hold.
    \item \textit{Charge/discharge rate} ($P$): Represents the amount of energy charged or discharged during one hour.
    \item \textit{Observation length} ($L$): Represents the number of recent prices observable to the agent.
\end{enumerate}

Each hour, the agent observes the following pieces of information:
\begin{enumerate}
    \item The $L$ most recent observable prices $\{p_{n-L+1},p_{n-L+2},\ldots,p_n\}$.
    \item The current charge $w_n$.
\end{enumerate}

Following the observation, the environment accepts one of three actions:
\begin{enumerate}
    \item \textit{Charge}: Increase the current charge by the charge/discharge rate or the maximum additional charge, whichever is lesser: $w_{n+1}=w_n+\min{(P,W-w_n)}$.
    \item \textit{Discharge}: Decrease the current charge by the charge/discharge rate or the current battery charge, whichever is lesser: $w_{n+1}=w_n-\min{(P,w_n)}$.
    \item \textit{Idle}: Do nothing: $w_{n+1}=w_n$.
\end{enumerate}

The battery's reward after taking an action is the resultant change in the value of its assets. ``Assets'' includes both monetary balance and the market value of any charge currently held. Note that, in this paradigm, no reward (positive or negative) can be incurred during the process of charging and discharging since both involve trading one asset for another of equal value. Instead, positive or negative rewards occur in the price changes between hours, which causes the charge held by the battery to change in value. Mathematically, the reward incurred during the transition to the $n^{th}$ time step is given as
\begin{align*}
    r_n = w_n\times\left(p_{n+1}-p_{n}\right).
\end{align*}

The agent's objective is to maximize cumulative reward over one year.

\section{Evaluation}

The simulation uses the DQN implementation provided by Stable Baselines3\footnote{https://stable-baselines3.readthedocs.io/en/master/modules/dqn.html}, a well-established Python RL library built on a PyTorch backend. Stable Baselines3's DQN implementation uses the default hyperparameters from \cite{MnihPlayingLearning}, the paper which introduced DQN, along with an Adam optimizer and learning rate of 0.0001. These parameters are left as-are for this simulation. Future work may improve performance by optimizing these parameters.

In this demonstration, the price signals are the actual hourly prices charged to customers enrolled in ComEd's hourly pricing program. These were collected directly from ComEd's website via their public API.
The battery is set to a capacity of $W = \SI{13.5}{\kilo\watt\hour}$ with a charge/discharge rate of $P=\SI{5}{\kilo\watt}$. These values are based on the Tesla Powerwall 2\footnote{https://www.tesla.com/powerwall} and give an approximation of the scale of storage system an Hourly Pricing customer might own. Finally, the observation length is chosen as $L = 48$, based on empirical analysis; that is, the performance was observed to improve as $L$ increased until around 48, after which improvement stopped.

\begin{figure}[htbp]
    \centerline{\includegraphics[scale=0.6]{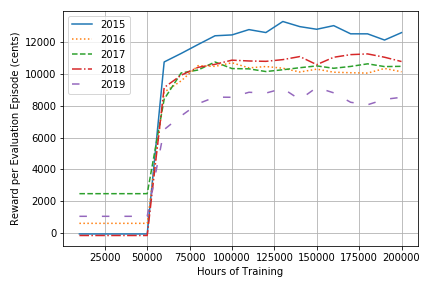}}
    \caption{Training curves for all five agents}
    \label{training}
\end{figure}

Five agents are trained, one for each year from 2015-2019 (2020 is avoided due to potential price irregularities during the COVID-19 pandemic).
During training, each agent begins with an arbitrary policy and plays through its training year price signal on a loop, updating its neural network every four time steps (hours). It trains on an epsilon-greedy policy for 200,000 total hours (about 22 one-year episodes), at which point improvement tends to be minimal. Every 10,000 hours, training is interrupted and the agent's performance is evaluated over a one-year episode using a \textit{greedy policy}, in which the agent always chooses the action with the highest Q-value. These periodic evaluations are called \textit{training curves} and are visualized in Figure \ref{training} for all five agents.
Note the difference in peak performance between the years, with some years providing substantially more opportunity for profit. This may be attributable to differences in weather, as high midday heat during the summer months has the potential to create higher and longer-lasting peak prices.

The models shows clear improvement during the learning process. The trained agents achieve profits of around \$100 per year, or around 27 cents per day - a reasonable figure, given the battery capacity, charging rate, and typical price variability (typically a couple cents per day).

The best-performing version of the agent (maximum point) from each training curve is chosen as the \textit{trained agent} from its respective year. Each trained agent is then cross-tested on the other four years. The purpose of testing on new data is to assess the extent to which the agent's policy, learned in one year, generalizes to other years. To account for differences in profit opportunity between the years, testing rewards are normalized to the same-year reward. For example, the normalized 2016 performance of the agent trained in 2015 is calculated as:
\begin{align*}
    \frac{\text{2015 agent's reward in 2016}}{\text{2016 agent's reward in 2016}}
\end{align*}

\begin{figure}[htbp]
    \centerline{\includegraphics[scale=0.6]{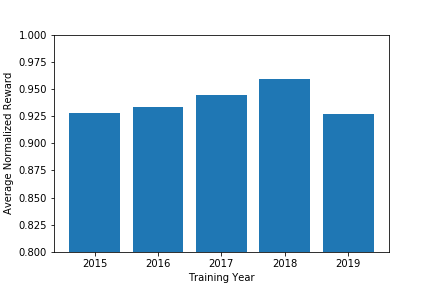}}
    \caption{Testing performance in non-training years}
    \label{testing}
\end{figure}

Figure \ref{testing} shows the average normalized reward in non-training years for all 5 agents.
The trained agents typically achieved around 94\% of the performance of their same-year counterparts, plus or minus around 2\%.

\begin{figure}[htbp]
    \centerline{\includegraphics[scale=0.6]{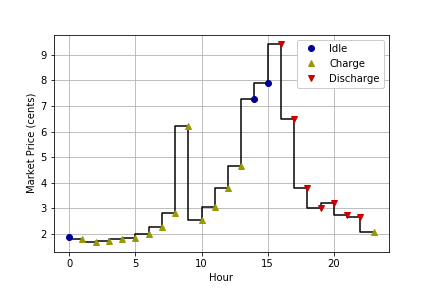}}
    \caption{Actions taken by the trained agent on a random day in 2018}
    \label{policy}
\end{figure}

Figure \ref{policy} qualitatively illustrates the learned policy. It gives the price signal of a random day in 2018 overlayed with the actions taken that day by the agent trained in 2018.  Note that charging or discharging more than three consecutive times is equivalent to idling, as the battery will have reached either an empty or full state. With this in mind, Figure 3 showcases the model's tendency to buy low and sell high.

\section{Summary and Conclusions}

This paper presented an application of RL for price-based demand response under an RTP policy. Five DQN agents, trained on the full ComEd Hourly Pricing curves from 2015-2019, were cross-evaluated. The preliminary results are encouraging for RL's prospects as a demand response tool in RTP environments. The training curves speak to the algorithm's capacity to identify and capitalize on patterns in the price signals, even given its limited information (the agent only observes price signals and charge levels). Furthermore, the relatively high testing performance indicates the policies were effective even when exposed to unseen data. The algorithm's performance may be improved in future work, perhaps by exposing the agent to more detailed observations.

\section{Future Work}

While the results indicate the potential viability of a model-free RL approach for residential demand response in an RTP environment, the work remains preliminary. Several simplifications present opportunities for further research:
\begin{enumerate}
    \item A real system would not be limited to making decisions only once per hour, and could switch states at any point within the hour. Incorporating this into the model would require smaller time steps.
    \item Batteries do not charge and discharge linearly. Instead, their charging/discharging rates are variable given the instantaneous level of charge.
    \item The model assumed 100\% round-trip efficiency, or the proportion of stored energy which is recoverable. Real batteries exhibit round-trip inefficiencies.
    \item Real batteries degrade over time. The damage associated with charging cycles could be represented in a similar problem as negative reward. Furthermore, there may be some \textit{switching cost}, or inefficiencies/damage incurred by changing modes of operation.
\end{enumerate}

While the agent appears to reach a plateau during training, this paper does not consider whether that plateau is optimal. This question could be addressed by attempting to outperform the trained agent using mathematical or model-based approaches. If the performance is indeed suboptimal, several tactics may achieve better results:
\begin{enumerate}
    \item Exposing the agent to more detailed observations, including ComEd's 5-minute prices.
    \item Switching to a modified DQN algorithm, such as Double DQN, Dueling DQN, or Rainbow DQN.
\end{enumerate}

In addition to improving performance in the simplified game presented in this paper, there are opportunities for extensions of the concept:
\begin{enumerate}
    \item Suppose the battery has some obligations other than turning a profit, such as keeping some power in reserve for blackouts. What if the battery is an electric vehicle with V2G capability, and must be ready to drive at appropriate times?
    \item A slightly modified version of the algorithm could model a home energy management system to control shiftable appliances such as water heaters and dishwashers instead of (or in addition to) batteries.
    \item Once optimal performance is achieved, how does a consumer's savings under RTP differ from their savings under other rate structures, such as flat-rate, TOU, and critical peak pricing?
    \item A utility-scale agent-based simulation with RTP and varying penetration of DQN-controlled distributed energy resources may shed light on the affect of widespread demand response on aggregate load curves. At some point, the market may become saturated so that the price signal flattens, thereby disincentivizing further demand response.
\end{enumerate}

\bibliographystyle{IEEEtran}
\bibliography{references}

\end{document}